\documentclass[10pt]{iopart}
\usepackage{amsthm}
\usepackage{amssymb}
\usepackage{graphicx}

\def\Th{\Theta}
\def\sig{\sigma}
\def\om{\omega}
\def\udot{\dot{u}}
\def\3nab{\tilde{\nabla}}

\def\hsp5{\hspace{5mm}}
\newcommand{\sfrac}[2]{{\textstyle{#1\over#2}}}
\def\case#1/#2{\textstyle\frac{#1}{#2}}

\def\be {\begin{equation}}
\def\ee {\end{equation}}
\def\bea {\begin{eqnarray}}
\def\eea {\end{eqnarray}}

\def\bc {\begin{center}}
\def\ec {\end{center}}
\newcommand{\hs}{\,-\,}

\def\case#1/#2{\textstyle\frac{#1}{#2} }

\def\jmp{{\it J. Math. Phys.}\ }
\def\pr{{\it Phys. Rev.}\ }

\def\cqg{{\it Class. Quantum Grav.}\ }

\def\ncim{{\it Il Nuovo Cim.}\ }

\def\prep{{\it Phys. Rep.}\ }

\def\jmp{{\it J. Math. Phys.}\ }

\begin{document}
%%%%%%%%%%%%%%%%%%%%%%%%%%%%%%%%%%%%%%%%%%%%%%%%%%%%%%%%%%%%%%
\title{Bounce Conditions in $f(R)$ Cosmologies}
%%%%%%%%%%%%%%%%%%%%%%%%%%%%%%%%%%%%%%%%%%%%%%%%%%%%%%%%%%%%%%
\author{Sante Carloni$^{1}$,  Peter K. S. Dunsby$^{1,2}$ 
and Deon Solomons $^{3}$}
\address{$1.$ Department of Mathematics and Applied Mathematics, \\
University of Cape Town, 7701 Rondebosch, South Africa.}
\address{$2.$ South African Astronomical Observatory, Observatory 7925, 
Cape Town, South Africa.}
\address{$3.$Cape Peninsula University of Technology, Cape Town, South Africa.}
%%%%%%%%%%%%%%%%%%%%%%%%%%%%%%%%%%%%%%%%%%%%%%%%%%%%%%%%%%%%%%
\begin{abstract}
%%%%%%%%%%%%%%%%%%%%%%%%%%%%%%%%%%%%%%%%%%%%%%%%%%%%%%%%%%%%%%
We investigate the conditions for a bounce to occur in
Friedmann\hs Robertson\hs Walker cosmologies for the class of
fourth order gravity theories. The general bounce criterion is
determined and constraints on the parameters of three specific
models are given in order to obtain bounces solutions. Furthermore,
unlike the case of General Relativity a bounce appears to be
possible in open and flat cosmologies.
\end{abstract}
%PACS number(s): 98.80.JK, 04.50.+h, 05.45.-a
%%%%%%%%%%%%%%%%%%%%%%%%%%%%%%%%%%%%%%%%%%%%%%%%%%%%%%%%%%%%%%%
\section{Introduction}
%%%%%%%%%%%%%%%%%%%%%%%%%%%%%%%%%%%%%%%%%%%%%%%%%%%%%%%%%%%%%%%
The idea of a ``bouncing'' universe in Friedmann\hs Robertson \hs
Walker (FRW) cosmologies has been examined many times since the
1930's when Tolman \cite{tolman} proposed that closed ($k=+1$) FRW
universes might re\hs expand after collapsing to a high
density state in the future \footnote{It was already known that
this was not possible for $k=0$ and $k=-1$ models.}. This subject
has remained popular throughout the history of cosmology
\cite{dipe} and is currently a very active area of research, 
motivated primarily by recent developments in $M$--theory, 
braneworlds and quantum gravity \cite{Mtheory}. 
In particular, the ekprotic model of the universe
\cite{ekprotic} is one such realisation of a cyclic cosmology
inspired by M\hs theory, while in loop quantum gravity the
semi\hs classical Friedmann equations have correction terms that
produce a bounce \cite{LQG}.

In classical cosmology such bounces are not possible in FRW models
if the active gravitational mass is positive: that is, if $\rho +
3p > 0$. This is a direct consequence of the Raychaudhuri equation
which is the fundamental equation of gravitational attraction
\cite{ray,ehl,ell}. On the other hand quantum fields and indeed
classical scalar fields $\phi$ can violate this condition
\cite{HE}. Hence, such fields can in principle allow bounce
behaviour in FRW models, however such bounces are difficult to
produce in universes that grow large enough to be realistic:
typically the probability of a bounce is of the order of the ratio
of the minimum to maximum expansion size (scale factor)
\cite{Starobinski,Barrow}. Furthermore very anomalous physical
behaviour can occur if these classical fields violate the {\it
reality condition}  $\dot{\phi}^2 \geq 0$, which is equivalent to
requiring that the inertial mass density is positive: $\rho + p
\geq 0$. It is also possible that ghost fields, i.e., fields that
have a negative energy density are capable of producing a
classical bounce \cite{ghost}. 

It is not only FRW models that are of interest. Smolin's idea of collapse 
to a black hole state resulting in re\hs expansion into a new expanding 
universe region \cite{smolin1,smolin2} suggests that the geometry 
of the universe at a bounce might be very different than the completely 
isotropic and spatially homogeneous FRW spacetimes; indeed, because some
spatially homogeneous modes are unstable \cite{we}, a more general
geometry might be expected. For example, in the scenario of black hole 
collapse and subsequent re\hs expansion, we might expect the geometry at the
bounce to be that of a Kantowski\hs Sachs model \cite{KS,ell67},
because this model has the same symmetries as the spatially
homogeneous interior region of the extended (vacuum) Kruskal
solution, that represents the late stage of evolution of an
isotropic black hole when the matter can be neglected.

Because of the growing interest in bouncing and cyclic cosmologies
it is worth exploring the full phase\hs space of possibilities
both in General Relativity and other models of gravity. Indeed
classical bounces in Kantowski\hs Sachs have recently been discussed in
\cite{KSbounce} within the context of standard General Relativity.

In this paper we examine the conditions for a classical bounce to
occur in models of gravity that have an ``effective" fourth order
action given by
\begin{equation}\label{lagr f(R)}
{\cal A}=\int d^{4}x\,{\cal L}=\int d^{4}x\,\sqrt{-g}\,
f(R)\;.
\label{Lagrangen}
\end{equation}
Our purpose is to use the 1+3 formalism in order to deduce the
condition for which fourth order gravity admits a bounce. Such
conditions will be given in terms of the parameter(s) for three
specific forms of $f(R)$: $R^{n},\exp(\lambda R)$.

In the last few years higher order theories have been proposed as
theoretical models for solving the problem of cosmological acceleration 
\cite{revnostra1,revnostra2}. The study of the bounce conditions for 
such theories allows us to determine if in fourth order gravity there 
is a connection between a bounce and the cosmological acceleration 
phenomenon. If this is the case, we would be able to use the bounce 
conditions to put constraints on cosmic acceleration and vice versa.

This paper is divided up in the following way: in section 2 we
give a brief overview of the fourth order theories of gravity 
$f(R)$; in section 3 we derive the bounce equations in the case 
of FRW cosmologies for a generic $f(R)$ theory; in sections 4--6 we
specialise the discussion to the three models listed above.
Finally in section 7 we present our conclusions.

Unless otherwise specified, natural units ($\hbar=c=k_{B}=8\pi
G=1$) will be used throughout the paper, Latin indices run from 0
to 3. The symbol $\nabla$ represents the usual covariant derivative 
and $\partial$ corresponds to partial differentiation. We use the
$(-,+,+,+)$ signature and the Riemann tensor is defined by
\begin{equation}
R^{\rho}{}_{\mu\lambda\nu}=W_{\mu\nu,\lambda}^{\rho}-W_{\mu\lambda,\nu}^{\rho}+
W_{\mu\nu}^{\alpha}W_{\lambda\alpha}^{\rho}+
W_{\mu\lambda}^{\beta}W_{\nu\beta}^{\rho}\;,
\end{equation}
where the $W_{\mu\nu}^{\rho}$ is the Christoffel symbol (i.e.
symmetric in the lower indices), defined by
\begin{equation}
W_{\mu\nu}{}^{\rho}=\frac{1}{2}g^{\rho\alpha}
\left(g_{\mu\alpha,\nu}+g_{\alpha\nu,\mu}-g_{\mu\nu,\alpha}\right)\;.
\end{equation}
The Ricci tensor is obtained by contracting the {\em first} and
the {\em third} indices
\begin{equation}\label{Ricci}
R_{ab}=g^{cd}R_{acbd}\;.
\end{equation}
Finally the Hilbert--Einstein action in presence of matter is defined by
\begin{equation}
{\cal A}=\int\sqrt{-g}\left[\frac{1}{2}R+ L_{m}\right]\;.
\end{equation}
%%%%%%%%%%%%%%%%%%%%%%%%%%%%%%%%%%%%%%%%%%%%%%%%%%%%%%%%%%%%%%%%%%%%%%%%
\section{Preliminaries} \label{Equazioni e convenzioni}
%%%%%%%%%%%%%%%%%%%%%%%%%%%%%%%%%%%%%%%%%%%%%%%%%%%%%%%%%%%%%%%%%%%%%%%%
By varying the action (\ref{Lagrangen}), we obtain the fourth order field 
equations
\begin{equation}\label{field eq f(R)}
f'(R)R_{ab}-\frac{1}{2}f(R)g_{ab}
=f'(R)^{cd}\left(g_{ca}g_{db}-
g_{cd}g_{ab}\right)+T^{M}_{ab}\;,
\end{equation}
where the prime denotes a derivative with respect to $R$. It is
easy to check that standard Einstein vacuum equations are
immediately recovered if $f(R)=R$.

In the 1+3 covariant formalism \cite{covariant}, after choosing 
a frame comoving with the average matter velocity vector $u^{a}$, 
the Raychaudhuri equation for a fourth order gravitational 
Lagrangian can be written as
\begin{eqnarray}\label{Ray}
\dot{\Th} - \3nab_{a}\udot^{a} =& -& \,\frac{1}{3}\,\Th^{2} +
(\udot_{a}\udot^{a}) - 2(\sig^{2} - \om^{2})-
\frac{1}{2f'}\,(\mu+3p) \nonumber \\&-& \frac{1}{2f'}\,(f-R
f'+3\ddot{f'}+\Th\dot{f'}-\triangle f') \ ,
\end{eqnarray}
and the Gauss\hs Codazzi equation (see equation (55) in
\cite{covariant}) is
\begin{equation}\label{Gauss-Cod}
\Th^{2} +3\frac{\dot{f'}}{f'}\Th  = \frac{3\,\mu}{f'} -
\frac{3}{2} \tilde{R} + 3\,\sig^{2}-\frac{3}{f'}\left[(f-R
f')+\triangle f'\right]\;,
\end{equation}
where $f=f(R)$, $f'(R)=df/dR$, $\tilde{R}$ is the 3--Ricci scalar,
$\triangle$ is the Laplacian operator and a ``dot" corresponds
to the covariant derivative along $u^{a}$. Note
that in this equation, as well as those that follow, we 
consider the Ricci scalar $R$ as an independent field. Such 
a position was first proposed in canonical quantisation of higher
order gravitational theories \cite{FRW nonlin} and allows us to write
the 1+3 equations as a system of second order differential
equations at the price of adding the constraint
\begin{equation}\label{R 1+3}
R=2\left(\dot{\Th}+\frac{2}{3}\Th^{2}+\sigma^{2}-2\omega^{2}
+\frac{\tilde{R}}{2}\right)\;.
\end{equation}
If we now consider the trace of (\ref{field eq f(R)})
\begin{equation}\label{traccia eq di campo}
3\ddot{f'}+3\Th\dot{f'}-3\triangle f'=f'R-2f+\mu-3p\;,
\end{equation}
we can simplify the Raychaudhuri equation to give
\begin{eqnarray}\label{Ray+trace}
\fl \dot{\Th} - \3nab_{a}\udot^{a} = - \,\sfrac{1}{3}\,\Th^{2} +
(\udot_{a}\udot^{a}) - 2(\sig^{2} - \om^{2}) - \frac{\mu}{f'} -
\frac{f}{2f'}+\frac{2}{f'}\left[\triangle f'-\Th\dot{f'}\right]\;.
\end{eqnarray}
Equations (\ref{Gauss-Cod}) and the (\ref{Ray+trace}) are the key
equations we need to determine the bounce conditions for this
class of cosmological models.

%%%%%%%%%%%%%%%%%%%%%%%%%%%%%%%%%%%%%%%%%%%%%%%%%%%%%%%%%%%%%%%%%%%%%%%%
\section{Bounce Equations} \label{bounce equation}
%%%%%%%%%%%%%%%%%%%%%%%%%%%%%%%%%%%%%%%%%%%%%%%%%%%%%%%%%%%%%%%%%%%%%%%%

In what follows, we define the occurrence of a bounce at time
$t=t_b$ by the conditions
\begin{equation}\label{bounce conditions}
\Th(t_{b})=0\;,~~~\dot{\Th}(t_b)>0\;.
\end{equation}
In FRW models \footnote{In anisotropic models which have more that
one scale factor $X_i$, $i=1,2,3$, the (\ref{bounce conditions})
has to be understood as characterising a bounce in the average
scale factor of the universe $S=\sqrt[3]{X_1X_2X_3}$. However one
can also consider a more generic situation where a bounce could
occur in any of the scale factors $X_i$. We can make this precise
by defining the expansion parameters $x_{i}=\dot{X_{i}}/{X_{i}}$,
so a bounce in $X_{i}$ will occur at time $t=t_b$ iff
$x_{i}(t_b)=0$ and $\dot{x}_{i}(t_b)>0$. It is clear that although
it may be possible to have a bounce in one of the scale factors
but not the other, this does not lead to a new expanding universe
region. We therefore require that a bounce occurs in all
$X_{i}$'s, even though they may in general occur at different
times.}, such conditions can be written as
\begin{equation}\label{bounce frw}
\dot{S}(t_{b})=0\;,~~~ \ddot{S}(t_b)>0\;.
\end{equation}
Let us now apply the bounce condition to equations
(\ref{Gauss-Cod}) and (\ref{Ray+trace}). In the case of FRW models
(where $\sigma=\omega=\dot{u}^a=0$) at $t=t_{b}$ we obtain
\begin{eqnarray}\label{bounce equation1}
\dot{\Th}_b= -\frac{\,\mu_b}{f'_b} +\frac{f_b}{2f'_b}\;,~~~
\tilde{R}_b  =
2\frac{\mu_b}{f'_b}+\frac{1}{f'_b}\left(f_b-R_bf'_b\right)\;,
\end{eqnarray}
and
\begin{equation}\label{Rbfried 1+3}
R=2\left(\dot{\Th}_b+\frac{\tilde{R_b}}{2}\right)\;,
\end{equation}
where we have used the suffix $b$ to indicate quantities evaluated
at the bounce time $t_b$. It is clear from the first of
(\ref{bounce equation1}) that the presence of the non--minimal
coupling affects the sign of $\dot{\Th}_b$ in two ways: through
the non--minimal coupling of higher order terms with matter and the
non--linear contribution to the Lagrangian. These two corrections
are related to each other, but in principle a bounce can be
induced by just one of them. The second equation leads to a surprising 
result: in higher order gravity a bounce does not necessarily
imply positive spatial curvature. This does not happen in 
standard GR, which can clearly be seen when $f(R)=R$.

Another interesting aspect of the system above, is that the second
equation in (\ref{bounce equation1}) is not independent of the
first one. This can be easily seen by substituting (\ref{Rbfried
1+3}) and then solving for $\dot{\Th}$. For this reason, we will 
consider only the first equation of (\ref{bounce equation1}) in our 
calculations:
\begin{equation}\label{bounce eqaution2}
\dot{\Th}_b= -\frac{\,\mu_b}{f'_b} +\frac{f_b}{2f'_b}\;.
\end{equation}
Substituting for the expansion in terms of the scale factor $S$
and using the standard result for the 3\hs curvature for FRW
models:
\begin{equation}
\Theta=3\dot{S}/S\;,~~~\tilde{R}=6K/S^2\;,
\end{equation}
equations (\ref{Rbfried 1+3}-\ref{bounce eqaution2}) can be
written as
\begin{eqnarray}\label{bounce equations FLRW}
\dot{\Th}_b= \frac{\ddot{S}_b}{S_b}=-\frac{\,\mu_b}{f'_b}
+\frac{f_b}{2f'_b}\;,~~~
R=6\left(\frac{\ddot{S}_b}{S_b}+\frac{K}{S_b^2}\right)\;.
\end{eqnarray}
Equations (\ref{bounce equations FLRW}) are the defining equations
for a bounce in FRW cosmologies and generalise the well known
results for General Relativity to $f(R)$ models. From now on, for
sake of simplicity, we will continue using the variables $(\dot{\Th}_b
,\tilde{R}_b)$, noting that the sign of $(\dot{\Th}_b
,\tilde{R}_b)$ correspond to the sign of $(\ddot{S}_b,K)$.

It is important to realise at this point, that since equations
(\ref{bounce equations FLRW}) are highly non--linear, we cannot
find an exact solution for them. However, since our purpose is
simply to investigate the possibility that a bounce may occur and 
not specific bounce models in fourth order gravity, we
can simply limit ourselves to checking the consistency of 
the system (\ref{bounce equations FLRW}) subject to the bounce 
conditions (\ref{bounce conditions}).

In the remaining sections of this paper we consider three specific
forms of $f(R)$: $R^{n}$, $R+\alpha R^{m}$ and $\exp(\lambda R)$.

%%%%%%%%%%%%%%%%%%%%%%%%%%%%%%%%%%%%%%%%%%%%%%%%%%%%%%%%%%%%%%%%%%%%%%%%
\section{$f(R)=R^{n}$} \label{caso R^{n}}
%%%%%%%%%%%%%%%%%%%%%%%%%%%%%%%%%%%%%%%%%%%%%%%%%%%%%%%%%%%%%%%%%%%%%%%%
In $R^{n}$\hs gravity, the function $f$ is specified by a generic power of 
the Ricci scalar: $f(R)=R^{n}$ \cite{revnostra1,revnostra2,santeRn}. In our 
analysis, the trivial cases $n=0$ and $n=1$ are neglected and $n$ may either
be a relative or a rational number.

In this case the bounce equation (\ref{bounce eqaution2}) is
\begin{equation}\label{bounce equations FLRW Rn1}
\dot{\Th}_b=\frac{R^n-2\mu}{2nR^{n-1}}\;,~~~
R=2\left(\dot{\Th}_b+\frac{\tilde{R_b}}{2}\right)\;,
\end{equation}
which can be written as
\begin{eqnarray}\label{bounce equations FLRW Rn2}
(n-1)\dot{\Th}_b  =
-\frac{\mu_b}{nR^{n-1}}+\frac{\tilde{R}_b}{2}\;,~~~
R=2\left(\dot{\Th}_b+\frac{\tilde{R}_b}{2}\right)\;.
\end{eqnarray}
In a previous paper, focused on the dynamics of $R^{n}$ gravity
\cite{santeRn}, it was shown that the sign of the Ricci
scalar remains unchanged once the initial conditions are fixed. 
This means that, if at the bounce we find the Ricci scalar to be positive, 
the sign of $R$ is positive for the entire cosmological history. Hence, we 
can consider the sign of $R$ as an ``initial condition''.

The form of (\ref{bounce equations FLRW Rn2}) suggests the
existence of two different behaviours depending on whether $n$ is 
bigger or smaller than 1 (see Table \ref{tavola Rn}). 
If $n>1$, the combination of (\ref{bounce equations FLRW Rn1}) and 
(\ref{Rbfried 1+3}) reveals that once the sign of $R$ and the set of 
values of $n$ is fixed, the sign of the 3--Ricci scalar is uniquely 
determined. In particular, a closed bounce occurs for most of the values 
of $n$ and $R$ with the only exception of $n$ even with $R <0$ and $n$
rational with odd denominator and even numerator with $R<0$.
\begin{table}
\centering
\caption{Sign of the 3--Ricci scalar for different type of values of $n$ in $ R^{n}$ 
gravity bounces. The numbers $r,q$ are chosen to be relative ($r,q\in{\mathcal Z}$).}
  \label{tavola Rn}
  \begin{tabular}{cccccc}
\br          $n$             & $n>1$& $n<1$   \\\mr
 $n\in {\mathcal Z}$ odd  & $\tilde{R}_b>0$ &  $\tilde{R}_b\lessgtr 0$\\
 $n\in {\mathcal Z}$ even & $\tilde{R}_b>0$ if $R>0$  &$\tilde{R}_b\lessgtr 0$ if $R>0$ \\
                          & $\tilde{R}_b<0$ if $R<0$ &  $\tilde{R}_b<0$ if $R<0$ & \\
$n=\frac{2r+1}{2q}$       & $\tilde{R}_b>0$ & $\tilde{R}_b\lessgtr 0$  \\
$n=\frac{2r}{2q+1}$       & $\tilde{R}_b>0$ if $R>0$  &$\tilde{R}_b\lessgtr 0$ if $R>0$ \\
                          & $\tilde{R}_b<0$ if $R<0$ &  $\tilde{R}_b<0$ if $R<0$ & \\
$n=\frac{2r+1}{2q+1}$     & $\tilde{R}_b>0$ & $\tilde{R}_b\lessgtr 0$ \\\br
     \end{tabular}
\end{table}
The same cannot be said for the case $n<1$.  In fact, with the only
exception of $n$ even with $R<0$ and $n$ rational with odd
denominator and even numerator with $R<0$, for which have 
an open bounce, the equation (\ref{bounce equations FLRW Rn2}) admits 
both an open and a closed bounce.

These results are very interesting if combined with those found in
the dynamical analysis of \cite{santeRn}. Here it was 
shown that a cosmological history exists which contains an
``almost--Friedmann" phase (AFP) followed by an accelerating
phase (ACP). Such a history is possible only for
\begin{eqnarray}\label{valori di n ok}
w=0,1/3 \qquad &\Rightarrow& \qquad\left\{\begin{array}{ccc}
n\lesssim 0.37 & \mbox{if}& n \in \mathcal{N}_{even}\;,\\
1.37\lesssim n\lesssim 2 & \mbox{if}& n \in \mathcal{N}_{odd}\;,
\end{array}\right. \\
w=1 \qquad &\Rightarrow&\qquad\left\{\begin{array}{ccc}
n\lesssim 0.37 & \mbox{if}  & n \in \mathcal{N}_{even}\;,\\
1.5\lesssim n\lesssim 2 & \mbox{if} & n \in \mathcal{N}_{odd}\;.
\end{array}\right.
\end{eqnarray}
where $\mathcal{N}_{even}$ is the set of even integers or rational
numbers with an even numerator, and $\mathcal{N}_{odd}$ is the set
of odd integers or rationals with odd numerator.

These values of $n$ can be traced back to our variables 
giving $R<0$ if $0<n<1$ and $R>0$ if $n<0$ and $n>1$. 

Comparing this with our results (beware of the change in the metric 
signature!) and bearing in mind that the sign of $R$ is fixed for a 
given cosmological history, we conclude that if $n>1$ and $n \in
\mathcal{N}_{odd}$, the presence of AFP$\rightarrow$ACP orbits is 
compatible with closed bounces for cosmic histories with $R>0$.
The same happens with $\mathcal{N}_{even}$ and $n<0$ with $R>0$ or
$0<n<1$ with $R<0$, where in the  second case $n$ can be only
rational.

Furthermore, the analysis of $R^{n}$--gravity with the supernovae
type Ia given in \cite{supernovae} shows us that $n$ must approximately
lie in the range $1.366<n<1.376$. This is within the limits given in equation
(\ref{valori di n ok}) and therefore the possibility of a closed bounce 
occurring in such cosmological histories is not excluded. 

%%%%%%%%%%%%%%%%%%%%%%%%%%%%%%%%%%%%%%%%%%%%%%%%%%%%%%%%%%%%%%%%%%%%%%%%
\section{$f(R)=R+\alpha R^{m}$} \label{caso R+R^m}
%%%%%%%%%%%%%%%%%%%%%%%%%%%%%%%%%%%%%%%%%%%%%%%%%%%%%%%%%%%%%%%%%%%%%%%%

This model has been one of the most popular fourth order
gravitational theories, and is still among those most studied.
There has been particular emphasis on the sub-case $m=2$, because 
these corrections have been shown to stabilise the divergence structure 
of gravity, allowing it to be renormalised (at least at first loop) 
\cite{renormalization}. In our study the parameter $\alpha$ will be 
taken to be real, but $m$ can either be relative or rational.

The general bounce equation (\ref{bounce eqaution2}) for
$f(R)=R+\alpha R^{m}$ reduces to
\begin{equation}\label{bounce equation R+Rm 1}
\dot{\Th}_b  = \frac{R_b+\alpha R_b^{m}-2\mu_b }{1+m\alpha
R_{b}^{m-1}}\;.
\end{equation}
Using the (\ref{Rbfried  1+3}) the same equation can be also given in
terms of the 3--Ricci scalar
\begin{equation}\label{bounce equation R+Rm 2}
 \tilde{R}_b=
 \frac{\alpha(m-1) R^{m}_b+2\mu_b }{1+m\alpha
R_{b}^{m-1}} \;.
\end{equation}
For our purposes {\it both} these versions of the bounce equation
are useful. In fact, from the first one we can obtain the values
of the Ricci scalar for which a bounce is possible (i.e.
$\dot{\Th}_b >0$) and then use them in (\ref{bounce equation R+Rm 2}) 
to obtain the sign of the spatial curvature. This is possible  because 
the $R_b$ term contains both $\dot{\Th}$ and $\tilde{R}_b$ 
and the two equations are non--linear in $R_b$, so that the 
system (\ref{bounce equation R+Rm 1}-\ref{bounce equation R+Rm 2}) 
may be considered as a parametric version of the full relation 
$\dot{\Th}_b(\tilde{R}_b)$. Unfortunately, the
complexity of the  (\ref{bounce equation R+Rm 1}) and (\ref{bounce
equation R+Rm 2}) makes it impossible to find  general exact results.
However, since most of the  important features of the
RHS of (\ref{bounce equation R+Rm 1}) and (\ref{bounce equation
R+Rm 2}) (number and sign of the solutions, etc.) depends only on
the nature of $m$, we can still give some general results.

A comparative analysis of the two equations above leads to the results 
listed in Tables 2--3, where the quantities $R^{*}_{i}$, 
$\bar{R}_i$, $R^{\odot}_{i}$ represent the values of $R$ for 
which $\dot{\Th}_b=0$, $\tilde{R}_b=0$, and $(1+m\alpha R_{b}^{m-1})\rightarrow 0$ 
respectively. We can see that closed bounces are allowed for every integer 
value of $m$ (often together with open bounces). For $m$ rational, closed bounces are
not allowed in general for $0<m<1$. For $m$ rational with even
denominator we have no closed bounce for $(m>1,\alpha<0)$  and no
bounce at all for negative $m$ and $\alpha$.
\begin{table}[t]
\label{tavola bounce R+aRm1} \caption{Sign of the 3--Ricci scalar
for $m$ integer. The quantities $R^{*}_{i}$, $\bar{R}_i$,
$R^{\odot}_{i}$ represent the values of $R$ for which
$\dot{\Th}_b=0$, $\tilde{R}_b=0$, and
$(1+m\alpha R_{b}^{m-1})\rightarrow 0$ respectively. The index
$i$, when present, indicates the presence of multiple roots. The
root with lower numerical value have lower index $i$ . Even if the
last two quantities can be calculate exactly in terms of the
values of the parameters, the first one requires numerical
calculation.} \centering
\bigskip
\begin{tabular}{cccccc}
\br &$(\alpha, m)$& $\dot{\Th}_b(R_b)>0$ & $\tilde{R}_b$\\\mr
$m\in{\mathcal Z}$ even &$m>0,\alpha>0$ & $R^{*}_{1}<R_b< R^{\odot}$       & $\tilde{R}_b<0$\\
                        &                &$R_b> R^{*}_{2}$                  & $\tilde{R}_b>0$\\
                        &$m<0,\alpha<0$  &$R^{\odot}<R_b<0$                 & $\tilde{R}_b<0$\\
                        &                &$R_b>0$                           & $\tilde{R}_b>0$\\
                        &$m>0,\alpha<0$  &$R^{\odot}<R_b< \bar{R}_1<0$      & $\tilde{R}_b<0$\\
                        &                &$R_b>\bar{R}_1$                   & $\tilde{R}_b>0$\\
                        &$m<0,\alpha>0$  &$R^{*}_1<R_b<0$                   & $\tilde{R}_b<0$\\
                        &                &$R^{*}_2<R_b<\bar{R}_1$           & $\tilde{R}_b>0$\\
                        &                &$\bar{R}_1<R_b<R^{\odot}$         & $\tilde{R}_b<0$\\
                        &                &$R_b>R^{*}_3>0$                   & $\tilde{R}_b>0$\\\mr
$m\in{\mathcal Z}$ odd  &$m>0,\alpha>0$  & $R_b> R^{*}$                     & $\tilde{R}_b>0$\\
                        &$m<0,\alpha<0$  &$R^{*}<R_b<0$                     & $\tilde{R}_b<0$\\
                        &                &$R_b>R^{*}>0$                     & $\tilde{R}_b>0$\\
                        &$m>0,\alpha<0$  &$R^{*}_{1}<R_b< R^{\odot}_1$      & $\tilde{R}_b<0$\\
                        &                &$R^{\odot}_2<R_b< \bar{R}$        & $\tilde{R}_b<0$\\
                        &$m<0,\alpha>0$  &$R_b> \bar{R}$                    & $\tilde{R}_b>0$\\\mr
$m=\frac{p}{2q}$        &$m>1,\alpha>0$  &$R_b> R^{*}$                      & $\tilde{R}_b>0$\\
$p,q\in{\mathcal Z}$    &$0<m<1,\alpha>0$&$R_b> R^{*}$                      & $\tilde{R}_b>0$\\
$p$ odd                 &$m<0,\alpha<0$  &no bounce                         &                \\
                        &$m>1,\alpha<0$  &$\forall R$                       & $\tilde{R}_b<0$\\
                        &$0<m<1,\alpha<0$&$0<R_b< R^{\odot}$                & $\tilde{R}_b>0$\\
                        &$m<0,\alpha>0$  &$R^{*}_{1}<R_b<\bar{R} $          & $\tilde{R}_b>0$\\
                        &                &$\bar{R}<R_b<R^{\odot} $          & $\tilde{R}_b<0$\\
                        &                &$R_b>R^{*}_2 $                    &$\tilde{R}_b>0$\\\br
\end{tabular}
\end{table}
\begin{table}
\label{tavola bounce R+aRm2}\caption{Sign of the 3--Ricci scalar
for $m$ rational. The quantities $R^{*}_{i}$, $\bar{R}_i$,
$R^{\odot}_{i}$ represent the values of $R$ for which
$\dot{\Th}_b=0$, $\tilde{R}_b=0$, and
$(1+m\alpha R_{b}^{m-1})\rightarrow 0$ respectively. The index
$i$, when present, indicates the existence of multiple roots. The
root with lower numerical value have lower index $i$ . Even if the
last two quantities can be calculate exactly in terms of the
values of the parameters, the first one requires numerical
calculation.} \centering
\bigskip
\begin{tabular}{cccccc}
\br &$(\alpha, m)$& $\dot{\Th}_b(R_b)>0$ & $\tilde{R}_b$\\\mr
$m=\frac{2r}{2q+1}$ \footnote{It is important to note that in this
case, even roots of the Ricci scalar appear in the action and the
field equations. This means that, before deriving the bounce
conditions, one should first determine that the cosmological
equations are meaningful at all times.}    &$m>1,\alpha>0$  & $R^{*}_1<R_b<0$                  & $\tilde{R}_b<0$\\
$r,q\in{\mathcal Z}$    &                & $R^{*}_2<R_b<0$                  & $\tilde{R}_b>0$\\
                        &$0<m<1,\alpha>0$&$R^{\odot}<R_b< 0$                & $\tilde{R}_b<0$ \\
                        &                &$R_b>R^{*}$                       & $\tilde{R}_b>0$ \\
                        &$m<0,\alpha<0$  &$R^{\odot}<R_b<R^{*}_{1}$         & $\tilde{R}_b<0$ \\
                        &                &$R_b>R^{*}_2$                     & $\tilde{R}_b>0$ \\
                        &$m>1,\alpha<0$  &$R^{\odot}<R_b< \bar{R}$          & $\tilde{R}_b<0$\\
                        &                & $R_b>\bar{R}$                    & $\tilde{R}_b>0$\\
                        &$0<m<1,\alpha<0$&$R_b>0$                           & $\tilde{R}_b<0$ \\
                        &$m<0,\alpha>0$  &$R^{*}_{1}<R_b<0 $               &$\tilde{R}_b<0$ \\
                        &                &$R^{\odot}<R_b< \bar{R}_2$        &$\tilde{R}_b>0$ \\
                        &                &$\bar{R}_2<R_b<R^{\odot}$         &$\tilde{R}_b<0$ \\
                        &                &$R_b>R^{*}_{3} $                  &$\tilde{R}_b>0$\\\mr
$m=\frac{2r+1}{2q+1}$   &$m>1,\alpha>0$  &$R_b>R^{*}_1$                     &$\tilde{R}_b>0$\\
$r,q\in{\mathcal Z}$    &$0<m<1,\alpha>0$&$R_b>R^{*}_1$                     &$\tilde{R}_b>0$\\
                        &$m<0,\alpha<0$  &$R^{*}_1<R_b<0$                   &$\tilde{R}_b<0$ \\
                        &                &$R_b>R^{*}_2$                     &$\tilde{R}_b>0$ \\
                        &$m>1,\alpha<0$  &$R^{*}_1<R_b<0$                   &$\tilde{R}_b<0$\\
                        &                &$0<R_b<  \bar{R}_1$               &$\tilde{R}_b<0$\\
                        &                &$R_b>\bar{R}_1$                   &$\tilde{R}_b>0$\\
                        &$0<m<1,\alpha<0$&$R_b>R^{*}_1$                     &$\tilde{R}_b<0$\\
                        &$m<0,\alpha>0$  &$R^{\odot}_1<R_b<0 $             &$\tilde{R}_b<0$\\
                        &                &$R^{\odot}_2<R_b<\bar{R}_1 $     &$\tilde{R}_b<0$\\
                        &                &$R_b>\bar{R}_1 $                 &$\tilde{R}_b>0$\\\br
\end{tabular}
\end{table}

%%%%%%%%%%%%%%%%%%%%%%%%%%%%%%%%%%%%%%%%%%%%%%%%%%%%%%%%%%%%%%%%%%%%%%%%
\section{Case $f(R)=\exp{(\lambda R)}$} \label{caso exp R}
%%%%%%%%%%%%%%%%%%%%%%%%%%%%%%%%%%%%%%%%%%%%%%%%%%%%%%%%%%%%%%%%%%%%%%%%

As a third example, we consider theories with a Lagrangians which may be 
expressed as an exponential of the Ricci scalar. This type of Lagrangian 
is interesting because it contains in some sense the previous two 
cases  due to the fact that the exponential can be developed 
in powers of the Ricci scalar. In other words, the study of an 
exponential Lagrangian gives us the chance to investigate, 
in a relatively  easy way, what happens if we consider 
a Lagrangian made up of a combination of different powers of the  Ricci scalar. 
In our analysis, the parameter $\lambda$ is taken to be an arbitrary  
real number.

The general bounce equation (\ref{bounce equation}) for an exponential  
Lagrangian becomes
\begin{eqnarray}\label{bounce equations expR}
\dot{\Th}_b  = \frac{1}{2\lambda}\exp(-\lambda
R_{b})\left[\exp(\lambda R_{b})-2\mu_b \right]\;.
\end{eqnarray}
This equation tells us that in order to have a bounce
($\dot{\Th}_b>0$), one of the two conditions below have to be
satisfied:
\begin{equation}
    \left\{
\begin{array}{l}
    \lambda>0,  \\
    R_b>\frac{\ln(2\mu_b)}{\lambda},  \\
\end{array}%
\right.\qquad
 \left\{%
\begin{array}{l}
    \lambda<0,  \\
    R_b<\frac{\ln(2\mu_b)}{\lambda}\;.  \\
\end{array}%
\right.
\end{equation}
Taking into account (\ref{R 1+3}), we see that only in the first
case a spatially closed bounce is possible. This result was
expected. In fact, if we perform a Taylor expansion of the
function $\exp{(\lambda R)}$ we have
\begin{equation}
    \exp(\lambda R)= 1+\lambda R+ \lambda^{2}R^{2}+....\;,
\end{equation}
so that in situations where the curvature is small, we obtain
\begin{equation}
    \exp(\lambda
R)\approx \lambda\left(\frac{1}{\lambda}+ R\right)\;,
\end{equation}
which is equivalent to the Hilbert Einstein action with a non--zero
cosmological constant. Consequently, $\lambda>0$ implies a positive
cosmological term in the action, which has been shown many times to
lead to a closed bounce.

%%%%%%%%%%%%%%%%%%%%%%%%%%%%%%%%%%%%%%%%%%%%%%%%%%%%%%%%%
\section{Conclusion}
%%%%%%%%%%%%%%%%%%%%%%%%%%%%%%%%%%%%%%%%%%%%%%%%%%%%%%%%%
In this paper we have derived the general bounce conditions for
Friedmann Robertson Walker models in fourth order gravity and
applied them to three specific models: $f(R)=R^{n}$,
$f(R)=R+\alpha R^{m}$ and $f(R)=\exp(\lambda R)$. 

Our analysis shows that in $R^{n}$--gravity a closed bounce occurs
for most values of $n$ and $R$ with the only exception being 
$n$ even with $R <0$ and $n$ rational with odd denominator and even
numerator with $R<0$. Comparing these results with the those obtained 
in \cite{santeRn}, we obtained the values of $n$ for which a cosmological 
history consisting of an almost--Friedmann phase followed
by an accelerated expansion is compatible with those that give
a closed bounce and found that they lie in the same interval compatible 
with observations of type Ia supernovae, given in \cite{supernovae}.

The $R+\alpha R^{m}$ models are more complicated due to the fact 
that there are two parameters involved. Closed bounces are allowed 
for every integer value of $m$ (often together with open bounces), 
but this is not true for rational values of $m$. In the 
case $m=2$, $\alpha>0$, our results are in agreement with the 
independent study done by Page \cite{Page:1987cb} using the 
Gibbons--Hawking--Stewart canonical measure.

For $\exp(\lambda R)$ models we found that a closed bounce occurs 
if $\lambda>0$ and the value of the Ricci scalar at the time of the
bounce is a particular function of energy density at the time of 
the bounce.

It is important to stress that the constraints found above
are necessary but not sufficient conditions and therefore this
paper only addresses the possibility of a bounce and not a specific
bouncing cosmological model. Nothing in our discussion implies
directly that the bounce occurs at an early stage of the
Universes' history\footnote{One could always argue that higher
order corrections become important when the curvature is high, 
typical of the very early evolution of the Universe, but 
in this case we assume that our {\em modification} to gravity 
holds in any curvature regime.}.

The most striking result of our study is that, contrary to what
happens in standard General Relativity, a bounce is possible in
cases in which the cosmology is not spatially closed. The idea of
a bounce in an open or flat universe may appear difficult to
visualise, but can be understood if we remember that the
quantity $\dot{\Th}_b$ gives a measure of the deviation of matter
worldlines. In this sense the bounce condition (\ref{bounce conditions}) 
simply means that there exists a phase in which the separation
between the matter worldlines decreases to a minimum and then 
increases again. Since this phenomenon is independent of the 
spatial geometry of the spacetime, the bounce itself is 
independent of it. On the other hand the existence of a $k<0$ 
bounce indicates that the problem of the cosmological bounce in 
fourth order gravity cannot always be related to the problem 
of gravitational collapse in such models. Indeed, if we compare 
the Friedmann equations in General Relativity
\begin{equation}
  \Theta=\pm\sqrt{3\rho-\frac{3\kappa}{a^{2}}}\;,
\end{equation}
and in Fourth Order Gravity
\begin{equation}\label{Gauss Cod f(r)}
\fl\Theta^{2}
-3\,\Theta\,\dot{R}\,\frac{f''(R)}{f'(R)}+\frac{3}{2f'(R)}\left[f(R)
-Rf'(R)\right]-\frac{3\mu_m}{f'(R)}-\frac{3\kappa}{S^{2}}=0\;,
\end{equation}
we realize that, because of the non--linearity of (\ref{Gauss Cod
f(r)}) many switches between contraction and expansion are in
principle possible. In other words, the cosmological history may be 
characterised by a sequence of contraction and expansion phases,
whose number and duration cannot be determined without an exact
solution of the cosmological equations. Such  ``wobbling" could in
principle be detected by its effect on structure formation and
CMB anisotropies, and in this way provide useful observational
constraints on higher order gravity theories.

How is it then possible to determine if a bounce ``$\acute{a}$ {\it la}
Tolman", separating two more or less causally disconnected phases
of the universe, is possible in fourth order gravity? A desirable 
constraint comes from the fact that we want such a bounce to be
an {\it absolute} minimum of the scale factor and that at this
minimum the energy density is higher than what it is during all 
the processes (recombination, etc) of classical cosmology, but lower 
than what it is when quantum gravity effects become important. 
Unfortunately, there seems to be no way to understand if the different 
bounce solutions found above satisfy these criteria without using an 
alternative methods or performing a direct analysis of the equations. 
Consequently, the existence of a cycling universe remains an open 
problem in fourth order gravity.

\ack This work was supported by the National Research Foundation
(South Africa) and the {\it Ministrero degli Affari Esteri- DG per
la Promozione e Cooperazione Culturale} (Italy) under the joint
Italy/South Africa science and technology agreement. We thank the
Italian group for their hospitality in the early stages of
development of the work.

%%%%%%%%%%%%%%%%%%%%%%%%%%%%%%%%%%%%%%%%%%%%%%%%%%%%%%%%%

\section*{References}

%%%%%%%%%%%%%%%%%%%%%%%%%%%%%%%%%%%%%%%%%%%%%%%%%%%%%%%%%

\end{document}